\documentstyle[12pt]{article}                  
\makeatletter
\def\nothing#1{}
\newdimen\earraycolsep
\renewcommand{\thetable}{\arabic{table}}
\renewcommand{\thefigure}{\arabic{figure}}
\def\author#1{{\pretolerance=10000 \raggedright \advance 
\leftskip by 1in \noindent #1 \vskip 1pc}}
\def\affiliation#1{{\advance\leftskip by 1in \noindent #1 \vskip -1pc}}
\def\refnote#1{{$^{\hbox{\scriptsize #1}}$}}
\def\affnote#1{\llap{$^{\hbox{\scriptsize #1}}$}}

\def\AmS{{\protect\the\textfont2%
        A\kern-.1667em\lower.5ex\hbox{M}\kern-.125emS}}

\def\p@LaTeX{{\family{times}\series{m}\shape{n}
   \selectfont L\kern-.36em\raise.3ex\hbox{\scriptsize A}
   \kern-.15em T\kern-.1667em\lower.7ex\hbox{E}\kern-.125emX}}

\newlength{\colwidth}

\def\@oddhead{\hfil}
\def\@evenhead{\hfil}
\def\fnum@figure{\footnotesize\raggedright{\bfseries \figurename~\thefigure.}}
\def\fnum@table{\normalsize\raggedright{\bfseries \tablename~\thetable.}}
\long\def\@makecaption#1#2{\vskip 10\p@ {#1 #2\par}}
\long\def\@makefntext#1{\setbox0=\hbox{$\m@th^{\@thefnmark}$}
    \noindent\hangindent=\wd0 \box0 #1}
\def\centerfig#1#2#3#4{\vspace*{#2}\relax
    \centerline{\hbox to#1{\special{#4:#3.#4 x=#1, y=#2}\hfil}}}
\newbox\@atbox
\long\def\atable#1#2#3{\begin{table}[tbp]\centering\footnotesize
\setbox\@atbox\hbox{#2}
\parbox{\wd\@atbox}{\caption{#1}}\par\smallskip
#2
\par\smallskip\parbox{\wd\@atbox}{\raggedright #3}
\end{table}}
\def\@nbibitem#1{\noindent \hangindent=2pc \hangafter=1
\refstepcounter{enumi}\hbox to 2pc{\arabic{enumi}.\hfil}%
\immediate\write\@auxout{\string\bibcite{#1}{\arabic{enumi}}}}
\makeatother 
\newcommand{\be}{\begin{equation}}
\newcommand{\ee}{\end{equation}}
\newcommand{\ba}{\begin{eqnarray}}
\newcommand{\ea}{\end{eqnarray}}
\newcommand{\pa}{\partial} 

\def\bb#1{\hbox{\mybb#1}}

\def\zet{\bb{Z}}

\def\R4{\bb{R}^4}
\font\mybb=msbm10 at 12pt

\def\C{{\cal C}}
\def\Hol{{\mbox{Hol}}}
\def\tr{{\mbox{Tr}}}
\def\Link{{\mbox{Link}}}
\def\sLink{{\mbox{sLink}}}

\def\unita{{1 \kern-.30em 1}}
\begin{document}
\begin{flushright}
{IFUM 544/FT}\\
{ROM2F/96/56}\\
\end{flushright}
\vskip 1cm
\leftline{NON LOCAL OBSERVABLES AND CONFINEMENT IN BF FORMULATION OF}
\leftline{YANG-MILLS THEORY\footnote{
Contribution to CARGESE SUMMER SCHOOL, july 96}
}
\vskip 1cm 

\author{Francesco Fucito,\refnote{1} \underline{Maurizio Martellini}, 
\refnote{2}~\refnote{3} 
and Mauro Zeni \refnote{2}}  

\affiliation{
\affnote{1}   
I.N.F.N. \ - \  Sezione di Roma II, \\
Via Della Ricerca Scientifica, \ \ 00133 \ Roma, \ \ ITALY \\ 
\affnote{2} 
Dipartimento di Fisica, Universit\`a di Milano and 
I.N.F.N.\ - \ Sezione di Milano, 
Via Celoria 16, \ \ 20133 \ Milano, \ \ ITALY \\
\affnote{3} 
Landau Network  at ``Centro Volta'', Como, ITALY\\
}
\begin{abstract}
The $vev$'s of the magnetic order-disorder operators in QCD are found 
with an explicit calculation using the 
first order formulation of Yang-Mills theory.
%
\end{abstract}

\setcounter{equation}{0}
\vskip 10mm 
\leftline{INTRODUCTION}
\vskip 4mm 
The definition of non local observables plays a fundamental role in the 
study of YM theory and QCD vacuum.Indeed the well known Wilson line operator
gives one of the most widely used confinement criteria,
namely the area law behaviour of its $vev$, associated with a linearly rising
confining potential between static probe charges in the QCD vacuum. 

The structure of the confining vacuum has been described by means of the 
condensation of ``magnetically'' charged degrees of freedom 
\cite{no,mandel,thsc,poly} giving
rise to the so-called  dual superconductor model. Several hypothesis on the
nature of these configurations and on the dynamical mechanism which leads
to their condensation have been formulated, and even if this 
picture in not yet
uniquely determined they are nevertheless believed to play a major role  in
the phenomenon of confinement. 

Actually this structure of the QCD vacuum  admits different 
possible phases 
\cite{thooft1}, not all confining, which can be labeled by means of 
the $vev$'s  of the Wilson loop $W$ 
and of an other non local operator $M$, the `t Hooft 
magnetic disorder parameter.  
$W$ and $M$ give rise to the well known `t Hooft algebra \cite{thooft1}, 
derived by the implicit definition of $M$ as producing center valued singular
gauge transformations along the magnetic vortex lines. 

The picture of the superconductor model for QCD is analytically realized 
in the first order formulation of YM theory, where an explicit definition 
of the color magnetic operator
$M$ is easily given and the calculation of the $vev$ of both $M$ and $W$
can be performed, displaying the expected behaviour for the confining phase. 

The first order form of pure Yang-Mills is described by the action 
functional \cite{fmz,ccgm,halp}
\be 
S_{BF-YM} = \int \tr (iB\wedge F +{g^2\over 4}B\wedge *B) 
\label{uno.1}
\ee 
where $F\equiv {1\over 2}F^a_{\mu\nu}T^a dx^\mu\wedge dx^\nu\equiv dA+
i[A,A]$, $D\equiv d +i[A,\cdot ]$ and $B$ is a Lie valued 2-form. 
The generators of the $SU(N)$ Lie 
algebra in the fundamental representation are normalized as 
$\tr T^aT^b=\delta_{ab}/2$ and the $*$ product (Hodge duality)
for a $p$ form in $d$ dimension is defined as 
$*=\epsilon^{i_1\ldots i_d}/(d-p)!$. 
The classical gauge invariance of (\ref{uno.1}) is given by 
$\delta A = D\Lambda_0$, $\delta B = -i[\Lambda_0 ,B]$  and 
the standard YM action is recovered performing path integration over $B$ or 
using the field equations 
\be 
 F ={ig^2\over 2}*B \qquad ,\qquad DB=0\quad .
\label{uno.2} 
\ee   
Off shell  $B$ does not satisfy a Bianchi identity and this fact should 
be related  with 
the introduction in the theory of magnetic vortex lines.
We remark that the short distance quantum behaviour of (\ref{uno.1})
is the same as in standard YM as it has been explicitely checked 
\cite{mz}. 

The action functional defined by the first term in the r.h.s. of 
(\ref{uno.1}) is known as the 4D pure bosonic BF-theory and defines a 
topological quantum field theory \cite{horouno}. 
Then the bosonic YM theory can be viewed as a {\it perturbative expansion} 
in the  coupling $g$ around the topological pure BF theory,  
the explicit topological symmetry breaking term $\sim g^2B^2$ in (\ref{uno.1}) 
introducing local degrees of freedom in the topological theory. 

The presence of the Lie-algebra valued two-form $B$ field in (\ref{uno.1})
allows the definition of an observable gauge invariant operator
\be                       
M(\Sigma ,C)\equiv \tr \exp \{ik
\int_{\Sigma } d^2y 
\ \Hol^y_{\bar x} (\gamma ) B(y) \Hol_y^{\bar x} (\gamma^{\prime} )\} \quad ,
 \label{tre.2} 
\ee 
where $\Hol_{\bar x}^y(\gamma )$ denotes the usual holonomy 
along the open path $\gamma\equiv\gamma_{\bar x y}$ with initial and final 
point $\bar x$ and $y$ respectively,
$\Hol_{\bar x}^y(\gamma )\equiv P\exp (i\int_{\bar x}^y dx^{\mu} A_{\mu}(x))$, 
where $k$ is an arbitrary expansion parameter, $\bar x$ is a {\it fixed} 
point  over the orientable surface $\Sigma\in M^4$ 
and the relation between the assigned paths 
$\gamma$, $\gamma^{\prime}$ over $\Sigma$ and the closed contour $C$ is the 
following: $C$ 
starts from the fixed point $\bar x$, connects a point $y\in\Sigma$ by 
the open path $\gamma_{\bar x y}$ and then returns back to the
neighborhood of $\bar x$ by 
$\gamma_{y\bar x }^{\prime}$, 
(which is not restricted to coincide with the inverse 
$(\gamma_{\bar x y})^{-1}=\gamma_{y\bar x }$). From the neighborhood 
of $\bar x$ the path starts again to connect another point 
$y^{\prime}\in\Sigma$. Then it 
returns back to the neighborhood of $\bar x$ and so on until all points
on $\Sigma$ are connected. The path 
$C_{\bar x}=\{ \gamma\cup\gamma^{\prime}\} $ is generic and we do not require 
any particular ordering prescription as it is done in similar constructions 
devoted to obtain a non abelian Stokes theorem \cite{ara}. 
Of course the 
quantity (\ref{tre.2}) is path dependent and our strategy is to regard it as 
a loop variable once the surface $\Sigma$ is given.

Using the hamiltonian formalism it is possible to show \cite{fmz} that 
$M(C)$  generates a local singular (or equivalently a multivalued
regular) gauge transformation, $\Omega_C (\vec x)$, along $C$: this 
is precisely  the defining 
property of the `t Hooft  color magnetic variable \cite{thooft1}. 
Using the classical constraints which arise from the action 
(\ref{uno.1}) it is possible to generate the classical gauge 
transformations; in order to 
have first class constraints the field content of the theory
has to be enlarged, 
including a Lie valued vector field $\eta$. This field corresponds to  
``topological" degrees of freedom which in our case become dynamical 
\cite{fmtz}. 

Given the classical algebra of tranformations, when switching 
to operator valued 
quantities, and considering the ordering procedures 
required by quantization,, 
one obtains (assuming $\Sigma\sim S^2$ and $k$ small) 
\be 
 M(C)|A> \simeq \tr \{ \unita+2ik\oint_C dy^i
\int_{\Sigma\sim S^2}d\sigma^{rs}_{(x)}\epsilon_{irs}
\delta^{(3)}(\vec y-\vec x)\delta^{ab}T_aT_b\} |A>\quad .
\label{tre.26}
\ee 
The integral in (\ref{tre.26}) can be expressed as 
\ba 
&&\oint_C dy^i
\int_{\Sigma}d\sigma^{rs}_{(x)}\epsilon_{irs}
\delta^{(3)}(\vec y-\vec x) =
-\oint_C dy^i
\int_{\Sigma}{d\sigma^{rs}_{(x)}\over 4\pi}\epsilon_{irs}
{\pa\over\pa x^l} \biggl[ {\pa\over\pa x_l} 
\biggl( {1\over |\vec y-\vec x|} 
\biggr) \biggr]
\nonumber        \\
  &\simeq & \lim_{\epsilon\to 0}{1\over 4\pi}\oint_C dy^i\oint_{C^{\prime}} 
dx^r\epsilon_{irl} 
{\pa\over\pa x_l}\biggl( {1\over |\vec y-\vec x|}\biggr) =
\Link(C,C^{\prime})_{\epsilon\to 0}\equiv\sLink(C)\quad ,
\label{tre.27} 
\ea 
where $\epsilon$ gives a point splitting regularization between $C$ and 
$C^{\prime}$, where 
$\Sigma=\Sigma^{\prime}\cup\Sigma^{\prime\prime}, \quad
\pa\Sigma^{\prime}=C^{\prime}$ and $C^{\prime}$ encloses the singularity of 
(\ref{tre.27}). 
The above linking number is the so-called  
self-linking number of $C$ \cite{rom}. 
While the linking number of two separated loops is not an invariant 
quantity in 4D, the quantity in (\ref{tre.27}) 
is well defined and finite, 
takes integer values and 
equals the number of windings of $C^{\prime}$ around $C$. 

Putting the above formulae in 
(\ref{tre.26}), we find 
\be 
 M(C)|A> \simeq \tr \{\unita ( 1+2ikc_2(t)\sLink(C) ) \} |A>\quad ,
\label{tre.30}  
\ee  
where $\delta^{ab} T_aT_b=c_2(t)\unita$. 
Eq. (\ref{tre.30}) implies that $M(C)$ generates an infinitesimal 
multivalued gauge transformation; 
whenever $C^{\prime}$ winds $n\equiv \sLink(C)$ times around $C$, 
$ M(C)$ 
creates a magnetic flux \cite{thooft1}
\ba 
\Phi_C\equiv 
{2kc_2(t)\over g}\sLink (C)\quad .
\label{tre.32} 
\ea 
The finite multivalued
gauge transformation $\Omega_C[\vec x]$ generated  by the action of 
$M(C)$ over some
state functional is given by
\be 
 M(C)|A(\vec x)>=|\Omega_C^{-1}[\vec x](A(\vec x)+id_{\vec  x})
\Omega_C[\vec x]>\simeq \tr \{ e^{ig\Phi_C}\unita\} |A(\vec x)> \quad .
\label{tre.33} 
\ee 
Because of the multi-valued nature of $\Omega_C[\vec x]$, 
since $A^{\Omega_C}\equiv \Omega_C^{-1}(A+d)\Omega_C$ should always be 
single valued, $\Omega_C$ must be in the center of $SU(N)$.
To recover the standard form of the center,
we normalize  the free expansion parameter as 
$k=2\pi/Nc_2(t)$.
With these normalizations the form of the color magnetic flux is given by
$\Phi_C=2\pi n/ Ng$ and the `t Hooft algebra is easily recovered 
\cite{thooft1}.

\vskip 10mm 
\leftline{COMPUTATION OF $<M(\Sigma , C)>$}
\setcounter{section}{2}
\setcounter{equation}{0}
\vskip 4mm 

In this section we compute the average of the BF-observable 
$M(\Sigma ,C)$ and precisely we consider 
the normalized connected expectation value 
$<M(C)>_{conn}\equiv {<M(C)> \over <1>}$.

In order to perform calculations we assume the scheme of the abelian projection 
gauge, in which $SU(N)$ is partially gauge fixed to an abelian subgroup 
\cite{thooft2}. 
In general this should be implemented using  an 
interpolating gauge \cite{thooft3}. 
Choosing (for example) 
$Y_{ij}={(F_{12})_{ij}/(\lambda_i-\lambda_j)},
\quad i\neq j=1,\ldots,N$, where the $\lambda$'s are the
eigenvalues of $F$, 
the proposed gauge condition is
$Y^{ch}+\xi D^0*A^{ch}=0$
where the superscripts $ch, 0$ stand for the off-diagonal and diagonal 
part of the matrix
$A=A^0+A^{ch}$ and $D^0$ is the covariant derivative with respect to the 
diagonal part of the gauge field. Interpolating gauges are such that for 
short
distances the relevant gauge is $ D^0*A^{ch}=0$ and the theory is 
renormalizable,
while for large distances the gauge is $Y^{ch}=0$. 
In standard YM theory we can find in the adjoint 
representation only a composite of the $F_{\mu\nu}$ (and its
covariant derivatives), 
but the dependence of $F_{\mu\nu}$ and $D^0*A^{ch}$ from the 
momentum $p_\mu$ 
is the same. 
Therefore the dominance argument doesn't apply. 
This is why the field $Y_{ij}$ was introduced.
Unfortunately this fact makes this 
gauge difficult to 
implement using  standard quantum YM fields.

Quite remarkably, these problems are not present in the BF-YM theory
due to the presence of the microscopic $B$ field. The 
interpolating gauge
can now be easily implemented choosing, for instance, 
$B^{ch}_{12}+D^0*A^{ch}=0$. 
Equivalently in our formalism we can diagonalize the two-form $B$ 
on the surface $\Sigma$, 
$\tilde B=V^{-1}BV=diag(\beta_1 ,...,\beta_N )$,  
$\beta_1\geq\beta_2 \geq\ldots\geq\beta_N $, 
with $V$ being the singular gauge transformation needed to implement the 
abelian gauge, 
and then use the background gauge condition $D^0*A^{ch}=0$ in the 
renormalization program. Indeed this is the way in which magnetic 
configurations, related to the abelian unbroken group, 
should enter the theory in the large distance regime and 
fill the vacuum, while in the same limit we may neglect 
the massive off-diagonal degrees of freedom $B^{ch}, A^{ch}$. 
This approximation
is often called Abelian dominance \cite{thooft2,eza}.

The existence of monopoles in the Abelian projection gauge is due
to the compactness of the $U(1)^{N-1}$ group and it is related to
the existence of non trivial topological objects for the entire $SU(N)$
theory. 
The reducibility of the $SU(2)$ gauge connection, implies that the
gauge bundle is split and thus
requires the existence of a positive definite first Pontrjagin class and 
intersection number (which we will define later on). 
This fact, in its turn, implies the absence of anti-self-dual harmonic
(closed) two forms \cite{freed}. 

In the general $SU(N)$ case
we now rewrite the functional integral in terms of the variables
$\alpha_i=[A^0]_{ii}$, $i=1,...,N$, $\sum_i\alpha_i=0$ and 
$f_i\equiv [F^0]_{ii} =d[A^0]_{ii}$. 
It is convenient to rescale the previous geometrical fields 
$A^0$, $B^0$ to the physical ones 
$A^0\rightarrow gA^0$,  
$B^0\rightarrow {1\over g}B^0$. 
Furthermore, we replace the surface integral in the definition of $M$ 
with the 
integration over the so-called Poincar\`e dual form, $\omega_{\Sigma}$, 
of (the homology class of ) the surface $\Sigma$ \cite{ccgm}.
By definition, $\omega_{\Sigma}$ is closed.
Moreover, choosing an orientation of the four manifold
and remembering the absence of anti-self-dual harmonic two forms 
(imposed by 
requiring the existence of a reducible gauge connection) 
$\omega_{\Sigma}$ is chosen to be self-dual {\it i.e.} 
$*\omega_{\Sigma}=\omega_{\Sigma}$ and  
with the property that (up to gauge one-forms) 
for a generic two-form $t$ 
\be 
\int_{\Sigma} t\simeq \int_{M^4}\omega_{\Sigma}\wedge t \quad .
\label{qua.10}
\ee  
In  a local system of coordinates  $(x,y,u,v)$ on $\Sigma$, so that $\Sigma$ 
is given by the equations $x=y=0$, 
the dual form $\omega_{\Sigma}$ can be taken as 
$\omega_{\Sigma}\simeq \delta^{(2)}(x,y)dx\wedge dy$ 
and normalized as 
\be 
\int_{N(\Sigma )}\omega_{\Sigma}\simeq \int \delta^{(2)} (x,y) dx\wedge dy =1
\quad ,  
\label{qua.11}
\ee 
where $N(\Sigma )$ is the transversal tubular neighbourhood on the surface 
$\Sigma$. 

Using the Abelian dominance approximation 
and defining the expansion parameter $k$ in units of the bare 
color charge $g$ with a suitable normalization, $k=2\pi qg$, 
the `t Hooft loop operator $M(C)$ becomes 
\ba 
&& M(C)={\rm Tr}\{ O_{ij}(C)\} \quad ,\label{qua.12} \\ 
&&O_{ij}(C)=\delta_{ij}\exp \{i2\pi q\int_{M^4}\omega_{\Sigma} \wedge
\beta_j [\cos (g\oint_C \alpha_j ) +i\sin (g\oint_C \alpha_j )]\} \quad ,
\nonumber 
\ea   
where the configurations $\alpha_i$ are to be intended as singular ones, 
related 
to the singularities occurring if two consecutive 
eigenvalues of $B$ coincide \cite{thooft2}. 
If $\beta_i =\beta_{i+1}$, we shall label such a point by $x^{(i)}$ and
$q_i$ will be the corresponding magnetic charge.
The key point will be the identification of 
the arbitrary expansion parameter $q$  with 
the magnetic charges $q_i$, i.e.  we will set 
$q_i\propto q\propto {1\over g}$. 
Consider then the magnetic order 
parameter $M$  in the $q\rightarrow 0$ limit, 
corresponding therefore at the strong coupling limit $g\rightarrow\infty$. 
In this limit, 
and relaxing the constraint $\sum_i \alpha_i =0$ 
thus extending the maximal torus 
of $SU(N)$ to $U(1)^N$ \cite{thooft4},
the operator (\ref{qua.12}) can be approximated by 
\be 
M(C)=\sum_i O_{ii}(C) \simeq 
N\exp \bigg\{{i2\pi q\over N}\int\omega_{\Sigma}\wedge
\sum_i\beta_i\cos (g\oint_C\alpha_i )\bigg\}\quad .
\label{qua.16}
\ee 

With all these approximations taken into account the $vev$ of the magnetic
order parameter in the strong coupling region becomes
\ba
&&<M(C)>_{conn}  
\simeq {N\over <1>}
\int {\cal D}\alpha_i {\cal D}\beta_i 
\exp \{-{1\over 2}\sum_i\int ({1\over 4}\beta_i\wedge *\beta_i 
+i\beta_i\wedge [d\alpha_i +{4\pi q\over N}\omega_{\Sigma}\cdot 
\nonumber   \\
&&\cdot\cos (g\oint_C\alpha_i )])\} = {N\over <1>}
\int {\cal D}\alpha_i  
\exp \{-{1\over 4}\sum_i\int  [d\alpha_i +{4\pi q\over N}
\omega_{\Sigma}\cos (g\oint_C\alpha_i )]^2)\}\ , 
\label{qua.17}
\ea
where the square of a form $t$ means $t\wedge * t$.
We now split the gauge fields as $\alpha_i =\bar\alpha_i +\hbar Q_i$, 
where  the quantum fluctuations $Q_i$ must be gauged (e.g. by a covariant 
gauge condition) 
and the $\bar\alpha_i$ are singular classical configurations.  
Postponing for a while the discussion of quantum fluctuations,
we concentrate on the semi-classical contribution to the path
integral which is given by 
\be 
<M(C)>_{conn}\simeq N\exp \{-{8\pi^2 q^2\over N^2}\sum_i
\int \omega_{\Sigma}\wedge\omega_{\Sigma^{\prime}}
\cos (g\oint_C\bar\alpha_i )
\cos (g\oint_{C^{\prime}}\bar\alpha_i^{\prime} )\}\quad ,
\label{qua.18b} 
\ee 
where 
$\Sigma^{\prime}$ represents a point splitting regularizations of $\Sigma$ 
for any point of $\Sigma$. 
The above equation has been obtained observing that the configurations 
$\bar\alpha_i$ obey the
(monopole) equation
\be 
*d\bar\alpha_i={4\pi q\over N}\omega_{\Sigma}
\cos (g\oint_C\bar\alpha_i)\quad ,
\label{qua.19}  
\ee
derived using 
the self-duality and closedness of $\omega_\Sigma$. Due 
to  the singular behaviour of the  $\bar\alpha_i$'s   
partial integration is not allowed \cite{kron} and no electric currents 
are present in the model. 
Equations of the type (\ref{qua.19})  appeared already
in the study of the duality properties of 
gauge theories and 4D manifold invariants \cite{fre,witmono}. 

The classical contribution to the action is
\be 
S_0= { 2\pi^2 m^2\over g^2}\int\sum_i
\omega_{\Sigma}\wedge\omega_{\Sigma^{\prime}}
\cos (g\oint_C\bar\alpha_i) 
\cos (g\oint_{C^{\prime}}\bar\alpha^{\prime}_i) \equiv 
{ 2\pi^2 m^2N\over g^2}Q(\Sigma,\Sigma^\prime)\quad ,
\label{qua.26}
\ee 
where Dirac quantization $g\oint_{C}\bar\alpha_i =2\pi n$ 
has been used, $m\in\zet$ labels monopole charges and 
$Q(\Sigma,\Sigma^{\prime})$ is the algebraic intersection number 
\cite{donald}. 
The closed curve $C^{\prime}\equiv \{ y^{\mu} (t)\} $ corresponds 
to the framing contour of $C\equiv \{ x^{\mu} (t)\} $, i.e. if it happens that 
$y^{\mu}(t)=x^{\mu}(t) +\epsilon n^{\mu}(t)$, with $\epsilon\to 0$, and 
$ |n^{\mu}|=1$ where $n^{\mu}(t)$ is a vector field orthogonal to $C$. 
Then $Q(\Sigma,\Sigma^{\prime})$ becomes the  
self-linking number of $C$,  
$Q(\Sigma,\Sigma^\prime)= \sLink(C)$. 
From a physical point of view, we may define $\sLink (C)$ as 
\be 
\sLink(C)\equiv {L(C)\over \rho}\quad ,
\label{qua.31}
\ee 
where $L(C)$ is the perimeter of the 
loop $C$ and $\rho$ plays the role of unit of 
lenght, reasonably associated to the vortex penetration depth.

Let us now discuss quantum fluctuations. If the effective theory for
large distances is a $U(1)$ type theory, for short distances the charged
degrees of freedom cannot be discarded anymore. Let $\Lambda$ be the scale 
separating these two regimes and let us divide the gauge field 
accordingly; 
moreover let, for the sake of simplicity, the gauge group be $SU(2)$. 
For scales bigger than $\Lambda$ we take $A^3=\bar\alpha+Q^3$ 
which is the 
usual $U(1)$ prescription. For scales smaller than $\Lambda$ we take 
$A^a=\delta^{a3}\bar\alpha+Q^a$, {\it i.e.} we continue the
classical solution into the small scales region where the quantum 
fluctuations coming from the charged degrees of freedom cannot be
discarded. The expectation is that the small scales behaviour is
insensitive to the classical solution according to the background field
method. Performing the functional integration over the quantum
fluctuations leads to a double contribution, in complete analogy
with the saddle point evaluation around an instanton background 
\cite{mrso}. 
The first contribution is given by a ratio of determinants given by 
\be 
R=\left[ {Det^\prime (-L_0)
\over Det^\prime(-L)}
\right] ^{1\over 2}\left[ {Det(-\bar D^2)\over Det(-\partial^2)}
\right]\quad ,
\label{qua.31bis}
\ee 
where $\bar D=d -ig\bar\alpha$, $\bar D^2\equiv\bar 
D^{\mu}\bar D_{\mu}$, 
$ L=\bar D^2\delta_{\mu\nu} -(1-{1\over\xi})\bar D_\mu\bar D_\nu
+2ig\bar F^0_{\mu\nu}(x)$, 
and $L_0$ is given from $L$ evaluated around the 
trivial background.
$\xi$ is the gauge parameter and in the primed determinants  
zero modes are omitted.  

The second contribution is given by the Pauli-Villars 
regularization 
of the determinants and it amounts to a scale $\mu$ 
(which is the 
Pauli-Villars mass) raised to a certain power which 
is given by 
the dimension of the moduli space of the 
classical solution.

Let us now proceed with the evaluation of these two 
contributions.
Using the self-duality property of our classical solution (\ref{qua.19}), 
the ratio
of determinants (\ref{qua.31bis}) can be written as
$R=\left[ Det(-\partial^2)/Det(-\bar D^2)\right]$ \cite{mrso}. This 
ratio has been
evaluated in Ref.\cite{cgot} using the heat kernel method 
in the case
of an $SU(N)$ gauge group but it is easy to specialize 
this result to our case. We obtain 
\be
\ln{Det(-\partial^2)\over Det(-\bar D^2)}=
{\ln({\mu\Lambda})\over 96\pi^2}
\int \bar F^0_{\mu\nu}(x) \bar F^0_{\mu\nu}(x)d^4x=
{\ln({\mu\Lambda})\over 96\pi^2}
{1\over 4}\int (d\bar\alpha )_{\mu\nu}(x)  
(d\bar\alpha )_{\mu\nu}(x)d^4x\quad .
\label{qua.35}
\ee 
The factor 1/4 comes from the normalization of the gauge group generators
according to the reduced connections $\bar\alpha_i$.

The contribution coming from the regularization of the zero modes 
is obtained
once the dimension of the moduli space is computed, according to 
Ref.\cite{witmono}, to be
\be
\dim {\cal M}=c_1(L)^2=c_1(L)\wedge c_1(L)={1\over 32\pi^2}\int 
(d\bar\alpha )_{\mu\nu}(x) (d\bar\alpha )_{\mu\nu}(x)d^4x\quad .
\label{qua.36}
\ee

Putting together the classical result (\ref{qua.26})
with the quantum fluctuations, 
we find that the bare coupling
$g$ can be substituted by its renormalized expression 
and that the exponent in (\ref{qua.18b}) can be written as
\be
{8\pi^2\over g^2_R}{c_1(L)^2\over 8}={c_1(L)^2
\over 8}\biggl({8\pi^2\over g^2}
-(8-{2\over 3})\ln ({\mu\Lambda})\biggr)
\equiv {c_1(L)^2
\over 8}\biggl({8\pi^2\over g^2}-\beta_1\ln ({\mu\Lambda})
\biggr)\quad ,
\label{qua.37}
\ee
where $\beta_1={22\over 3}$ is the first coefficient of the $SU(2)$ 
beta function of the non-abelian Yang-Mills theory.

\vskip 10mm 
\leftline{THE AVERAGE OF THE WILSON LOOP }
\setcounter{section}{3}
\setcounter{equation}{0}
\vskip 4mm 

In this section we shall compute the average of the Wilson loop
and find an area law behaviour for its leading part. 
Furthermore, in our formalism, the area law gets a nice geometrical 
interpretation: it is the response of the true QCD vacuum to arbitrary 
deformations of the quark loop ${\cal C}$.

The starting point here is given by the Wilson loop operator written in 
terms of the non abelian Stokes theorem (see e.g. \cite{ara} ):
\be
W_t(\C) \equiv W_t(\Sigma ,C)=Tr_t\{ P_{\Sigma}
\exp [i\int_{\Sigma} \Hol_{\bar x}^x (\gamma )
F(x) \Hol^{\bar x}_x(\gamma^{\prime})]\}\quad ,
\label{cin.1} 
\ee
where $\C=\partial\Sigma$, $C=\{\gamma(x)\cup\gamma^{\prime}(x)\}$ 
was defined at
the beginning of section 2 and $P_{\Sigma}$ means surface path ordering.
$W_t$ is 
calculated with respect to some irreducible representation $t$ of $SU(N)$. 
\be 
<W_t(\C)>_{conn}\equiv {<W_t(\Sigma ,C )>\over <1>}
\equiv 
{\int {\cal D}B{\cal D}A \ W_t(\Sigma ,C )e^{-S_{BF-YM}}\over 
\int {\cal D}B{\cal D}A\ e^{-S_{BF-YM}}} \quad ,
\label{cin.2}
\ee 
where $S_{BF-YM}$ was defined in (\ref{uno.1}). 
Expanding in series the Wilson loop (\ref{cin.1}) we get 
\be 
<W_t(\Sigma,\C)>=\sum_n<{1\over n!}{\tr}_t 
P_\Sigma\int_{\Sigma_1}\ldots\int_{\Sigma_n}
(i\Hol(\gamma )F(x) \Hol^{-1}(\gamma^{\prime}))^n>\quad .
\label{cin.4}
\ee
We then use the identity 
\be  
e^{-\frac i 4 \int *B^a_{\mu\nu}F^a_{\mu\nu}} F^a_{\rho\sigma}(x)=
4i\frac {\delta} {\delta *B^a_{\rho\sigma}(x)}(e^{-\frac i 4 \int 
*B^a_{\mu\nu}F^a_{\mu\nu}})\quad ,
\label{cin.6}
\ee  
and performing a partial integration with respect to the functional
derivative in (\ref{cin.6}) we can replace, in the path integral,
\be 
i\Hol(\gamma )e^{-\frac i 4 \int *B^a\cdot F^a}F(x) 
\Hol^{-1}(\gamma^{\prime})\rightarrow 
4\Hol(\gamma )e^{-\frac i 4 \int *B^a\cdot F^a}
({\delta\over \delta *B(x)})\Hol^{-1}(\gamma^{\prime})\quad .
\label{cin.7}
\ee 
The functional derivative acts only to its right on the exponential
of the mass term $-g^2/16\int B^a_{\mu\nu}B^a_{\mu\nu}$, since $\Hol$ does not
contain the $B$ field.
We now need the identity 
\be 
V(\frac {\delta} {\delta *B^a(x)_{\rho\sigma}})e^{-\frac {g^2} {16}
\int B^a_{\mu\nu}  
B^a_{\mu\nu}}=
V(-\frac {g^2} {8} *B^a_{\rho\sigma})e^{-\frac {g^2} {16}\int B^a_{\mu\nu}
B^a_{\mu\nu}}\quad ,
\label{cin.10}
\ee 
where $V$ is the functional defined by 
\be 
V(\frac {\delta} {\delta *B^a(x)})\equiv 
P_\Sigma\int_{\Sigma_1}\ldots\int_{\Sigma_n}
(4\Hol(\gamma ) 
(\frac {\delta} {\delta *B^a(x)}) 
\Hol^{-1}(\gamma^{\prime}))^n\quad .
\label{cin.11}
\ee 
Resumming the exponential series for the Wilson loop 
we finally get the ``duality" relation 
\be 
<W_t(\C)>_{conn}={<M^*_t(\Sigma ,C,\C=\partial\Sigma)>
\over <1>}\quad , 
\label{cin.9}
\ee 
where
\be 
M^*_t(\Sigma ,C,\C=\pa\Sigma)=
Tr_t[P_{\Sigma} \exp \{ -{g^2\over 2}\int_{\Sigma}
\Hol_{\bar x}^x (\gamma )
*B(x) \Hol^{\bar x}_x(\gamma^{\prime})\} ] \quad .
\label{cin.8}
\ee 
$M^*_t(\Sigma ,C)$ is the dual (in the sense that 
$B\to *B$) of the observable 
$M_t(\Sigma ,C)$ defined in (\ref{tre.2}) with $k$ set to $k=ig^2/2$.

To calculate (\ref{cin.9}) we expand perturbatively  in $g$ both 
the exponential and 
the holonomies which appear in the exponent of (\ref{cin.8}) \cite{cm}. 
The first relevant contraction 
encountered at lower level is given in terms of $<A*B>$, 
which can be computed starting from the off-diagonal 
propagator $<AB>$ \cite{mz}. Therefore we find 
\be 
<M^*_t(\C)>_{conn}  =e^{-{g^2\over 2} c_2(t)\oint_{C}\int_{\Sigma}<A*B>}
\Delta (\Sigma)\quad .
\label{cin.13}   
\ee 
$\Delta (\Sigma)$ depends 
on higher order integrations over $\Sigma$ which do not simply involve the  
quantity $\oint_{C}\int_{\Sigma}<A*B>$ alone; therefore while 
the explicit calculation of $\Delta (\Sigma )$ is an open 
problem, for the purpose of showing the area law behaviour
its knowledge should not be essential.   
Consider now the integral in the exponent of (\ref{cin.13}), 
\ba 
&& \oint_C\int_{\Sigma}<A*B> = 
<\oint_C A\int_{\Sigma}d\sigma^{\mu\nu}(x)(*B(x))_{\mu\nu} >\\
\nonumber 
&& =<\oint_C A\int_{\Sigma}(*d\sigma )^{\mu\nu}(x) B(x)_{\mu\nu} > =
\oint_C dy^{\lambda}\int_{\Sigma}(*d\sigma)^{\mu\nu}(x) <A(y)_{\lambda}
B(x))_{\mu\nu} >\ ,
\label{cin.29}
\ea 
where 
$*d\sigma (x)$ is the infinitesimal surface element of the plane $\Sigma_x^*$ 
dual to $\Sigma$ at the point $x\in\Sigma$. We may rewrite 
\be 
\oint_C dy^{\lambda}(*d\sigma)^{\mu\nu}(x) <A(y)_{\lambda}B(x))_{\mu\nu}>=
\oint_C dy^{\lambda}\int_{\Sigma^*_x}\omega_{\Sigma}^{\mu\nu} 
<A_{\lambda}B_{\mu\nu} > \quad ,
\label{cin.30}
\ee 
recalling that $\omega_{\Sigma}$ is locally given by a bump function with 
support on $\Sigma$. Eq. (\ref{cin.30}) is by definition the linking number 
between the curve $C$ and the dual plan $\Sigma^*_x$ in $x$ to $\Sigma$.  
Indeed the linking $\Link(C,\Sigma )$, whith arbitrary  $C$ and $\Sigma$, is 
defined by \cite{horo} 
\ba 
\Link(C,\Sigma ) &=&
\frac 1 {8\pi^2}\oint_{C}dx^{\alpha}
\int_{\Sigma}d\sigma^{\beta\gamma}(y)\epsilon_{\alpha\beta\gamma\delta}
{(x-y)^{\delta}\over |x-y|^4} \nonumber \\ 
&=& 4\oint_{C}dx^{\alpha}
\int_{\Sigma}d\sigma^{\beta\gamma}(y)
<A_{\alpha}^a(x)B_{\beta\gamma}^a(y)>  \quad .
\label{cin.12}   
\ea   
In our case, by construction, $\Link(C,\Sigma^*_x)\neq 0$. The residual 
integration over $\Sigma$ in (\ref{cin.29}) spans all the dual $\Sigma^*$ 
to $\Sigma$, yielding a contribution proportional to the area of $\Sigma$, 
\ba  
<W_t(C)>_{conn} &\sim & \exp \{ -{g^2c_2(t)\over 8}
\int_{\Sigma} \Link(C,\Sigma^*_x)\} \quad , \nonumber \\ 
\int_{\Sigma} \Link (C,\Sigma^*_x) &\sim & A(\Sigma ) \quad . 
\label{cin.17}
\ea 
We may get a better understanding of (\ref{cin.17}) considering a lattice 
regularization of $\Sigma$, i.e. $\Sigma\to\Sigma_{PL}$. In this case $C$ 
runs over the links of the lattice, while $\Sigma_x^*$ corresponds to 
an element of the dual lattice through the plaquette centered at $x$. 
$\int_{\Sigma_{PL}}\Link(C,\Sigma^*_x)$ is an integer which 
counts the number $N_v$ of vertices 
of the dual lattice on $\Sigma_{PL}$ or equivalently the number of plaquettes 
$N_P$  of $\Sigma_{PL}$. We may then write
\be 
\int_{\Sigma_{PL}}\Link(C,\Sigma^*_x)= N_P={A(\Sigma_{PL})\over a^2}\quad ,
\label{cin.18} 
\ee  
where $A(\Sigma_{PL})$ is the minimal area 
bounded by the quark loop $\C$ and $a$ is 
the lattice spacing. Up to this point the quantities which enter the 
calculations are the bare ones, but when passing to the continuum limit  
$a\to 0$ the renormalization of the theory implies that they are replaced by 
the renormalized ones. In particular 
$N_P =A(\Sigma_{PL})/a^2$ should become $A(\Sigma)/l^2$ with $l$ a typical
scale in QCD which, owing to a proper 
dressing of the off diagonal propagator entering (\ref{cin.13}), 
should be choosen as $\Lambda_{QCD}^{-1}$. 
Therefore one may rewrite (\ref{cin.17}) as 
\be 
<W_t(\C)>_{conn}\sim \exp [-\sigma(\Lambda_{QCD}) A(\Sigma)]\quad ,
\label{cin.19}
\ee 
where $\sigma(\Lambda_{QCD})$ is the string tension defined by 
\be 
\sigma(\Lambda_{QCD}) \equiv {g^2_R c_2(t)\over 8}
\Lambda^2_{QCD}\quad 
\label{cin.19bis} 
\ee 
and where we have replaced the bare coupling constant $g$ with the 
running coupling constant $g_R$. The way in which the continuum limit is 
reached and the fields are dressed is presently not under our 
control but the key geometrical 
feature of the area law should remain unchanged. 

A deeper investigation should also clarify some other questions. While the 
area law displayed in (\ref{cin.19}) corresponds to the expected confining 
behaviour of QCD vacuum, no screening is shown when the test charges are taken 
in the adjoint representation and the behaviour in this case is closer 
to what expected in the large N limit \cite{make}. 
It might seem puzzling that the area law behaviour comes out of a perturbative 
calculation. In reality our perturbative expansion has to be understood as 
an expansion around the gauge fixed topological BF theory 
whose vacuum contains ``relevant'' topologically non-trivial configurations. 
It is thus resonable to imagine that the non perturbative information comes out 
of this non-trivial vacuum structure. 
The contribution of these ``magnetically charged'' configurations is expected 
to give rise to a mass gap in the model and we guess a possible form for it,
namely 
\be 
m^2\sim \mu^2 e^{-\frac{2\pi}{q^2}}\quad ,
\ee 
where  $\mu$ is a regulator mass and $q$ is the magnetic charge satisfying a 
proper Dirac quantization $qg=2\pi n$. A suitable dressing displaying such a 
pole for propagators is obtained substituting (in euclidean momentum space) 
the $1/p^2$ denominators with $1/(p^2 +\Pi (p^2))$, where 
$\Pi (p^2)\sim 2(\frac{q^2}{4\pi})p^2\ln (\frac {p^2}{\mu^2})$. In such a way, 
in a $q\to 0$ limit (i.e. strong coupling expansion) and applying 
Dirac quantization, the string tension read 
out of (\ref{cin.17}) should be vanishing when the adjoint representation is 
choosen. 

A last point to be discussed is the apparent 
surface dependence which the Wilson 
loop acquires in the representation (\ref{cin.9}). Deforming the surface 
bordered by the loop $\C$ corresponds to a ``gauge'' transformation on the 
connections defined on the loop space; in this sense the ``gauge'' 
independence of 
observables in loop space 
corresponds to surface independence for the Wilson line. 
Lastly, our formalism appears quite similar to that recently 
introduced in \cite{poly} for the $vev$ of $W$.

M.M. acknowledges the hospitality of the Board of the Cargese School. 
This work has been partially supported by MURST. 
M.~Z. is associated to the University of Milan according to
TMR programme  ERB-4061-PL-95-0789.
 

\end{document}